# Plasma Processing of Large Curved Surfaces for SRF Cavity Modification

J. Upadhyay,[1] Do Im,[1] S. Popović,[1] A.-M. Valente-Feliciano,[2] L. Phillips,[2] and L. Vušković[1]

[1]Department of Physics - Center for Accelerator Science, Old Dominion University, Norfolk, VA 23529, USA

[2]Thomas Jefferson National Accelerator Facility, Newport News, VA 23606, USA

Plasma based surface modification of niobium is a promising alternative to wet etching of superconducting radio frequency (SRF) cavities. The development of the technology based on $Cl_2$/Ar plasma etching has to address several crucial parameters which influence the etching rate and surface roughness, and eventually, determine cavity performance. This includes dependence of the process on the frequency of the RF generator, gas pressure, power level, the driven (inner) electrode configuration, and the chlorine concentration in the gas mixture during plasma processing. To demonstrate surface layer removal in the asymmetric non-planar geometry, we are using a simple cylindrical cavity with 8 ports symmetrically distributed over the cylinder. The ports are used for diagnosing the plasma parameters and as holders for the samples to be etched. The etching rate is highly correlated with the shape of the inner electrode, radio-frequency (RF) circuit elements, chlorine concentration in the $Cl_2$/Ar gas mixtures, residence time of reactive species and temperature of the cavity. Using cylindrical electrodes with variable radius, large-surface ring-shaped samples and d.c. bias implementation in the external circuit we have demonstrated substantial average etching rates and outlined the possibility to optimize plasma properties with respect to maximum surface processing effect.

## I. INTRODUCTION

To improve the RF performance of SRF niobium cavities, the cavity surface must be prepared by a process that enhances surface smoothness, removes impurities and creates less sharp grain boundaries. Currently used technologies are buffered chemical polishing or electro polishing [1]. These technologies are based on the use of hydrogen fluoride (HF) in liquid acid baths, which poses major environmental and personal safety concerns. HF-free plasma-based ("dry") technologies are a viable alternative to wet acid technologies as they are much more controllable, less expensive, and more environment-friendly. To the best of our knowledge, we are presenting the first result for three dimensional plasma etching of the inner surface of bulk niobium cavities.

As a proof of concept, we have developed in earlier work an experimental setup for etching small flat niobium samples [2, 3]. Microwave plasma at 2.45 GHz, inside a quartz tube, was used for this experiment. The gas mixture consisted of up to 3% chlorine diluted in argon. The results with flat samples were very encouraging [3], with etching rates up to 1.7 μm/min. In every case, the surface roughness of plasma etched samples were equal or lower than the chemically etched samples.

The next line of action in the development of plasma etching technology for SRF cavities is to perform the removal of the niobium from the inner surface of the cylindrical cavity. This step signifies the transition from flat coupons to large curved surfaces before we proceed to etch the SRF cavity. To achieve plasma etching in cylindrical geometry, this also means that the transition has to be made from symmetric to asymmetric discharges in the coaxial electrode arrangement, since the inner and the outer electrode do not have the same surface area.

In order to correlate the parameters of the cylindrical asymmetric discharge and the etching rate, we have introduced a curved ring niobium sample with a surface area substantially larger than flat coupons [4]. We have achieved plasma etching by a cylindrical coaxial capacitively coupled RF discharge, the properties of which are described in the following Section. The actual experimental set-up for our present work is described in the third Section, and the experimental results are discussed in the fourth Section.

## II. COAXIAL CAPACITIVELY COUPLED RF PLASMA

In the experiments described here the coaxial plasma is generated using an electrode running coaxial with the material cylinder to be polished. The coaxial plasma is excited with an RF wave form of 13.56 MHz. This frequency was chosen because its wavelength is much larger than the characteristic size of the sample being polished, and is expected to give a more uniform polish.

The plasma produced is in the capacitively coupled plasma (CCP) regime. In CCP when one electrode has a smaller area than the other, to maintain the current continuity, the smaller area electrode acquired a negative d.c. voltage called self-bias [5, 6] as shown in Fig.1.

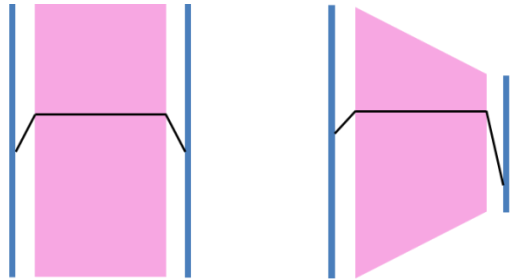

FIG. 1. Plasma sheath potential distribution in symmetric and asymmetric plasma.

The properties of the asymmetric discharges were extensively studied in the context of the development of planar semiconductor processing technology starting as early as the 1970s [6, 7]. Koenig and Maissel [7] used simple arguments, such as constant ion density near both electrodes, Child's law and absence of collisions in the sheaths, to arrive at the conclusion that the inverse scaling of electrode's voltage and surface area follows a power law with the scaling exponent of four. A number of later experimental works implied a much lower exponent [8 – 11]. We are presently interested in the cylindrical geometry in particular, where some experiments and models do exist [10, 12].

The shape of an SRF cavity presents a challenge for the RF plasma processing of its inner wall. It has a curved cylindrical symmetry and therefore, the processed surface has a larger area than the surface of the inner electrode. By contrast, the mature technology, such as semiconductor wafer processing, is based on planar geometry. Moreover, the wafer to be etched is placed on the smaller-area electrode in order to take advantage of the asymmetry in the plasma sheath voltage. In the present case, the cavity that has to be etched is grounded and has a large surface area, so, in the absence of positive d.c. voltage on the inner electrode, the sheath potential is substantially lower on the cavity surface than the sheath potential of the inner electrode.

Although we produce chemical radicals (excited neutral atom, molecule and ions) of chlorine by the plasma to carry on the required reactions for material removal from the niobium surface, we need a potential for the ions to be accelerated prior to hitting the surface. Therefore, a grounded niobium surface etching requires bringing the driven (inner) electrode to a positive d.c. potential with an RF power. Details about the self-bias potential and the dependence of etch rate on other parameters explaining the etch mechanism will be published in a separate paper.

## III. EXPERIMENTAL SETUP

We developed an experiment with a simple cylindrical cavity of 7.2 cm internal diameter and 15 cm length. This cylindrical cavity has 8 mini conflat ports. Some serve for holding niobium samples and some had a view port for diagnostic purposes. The cavity is shown in the Fig. 2.

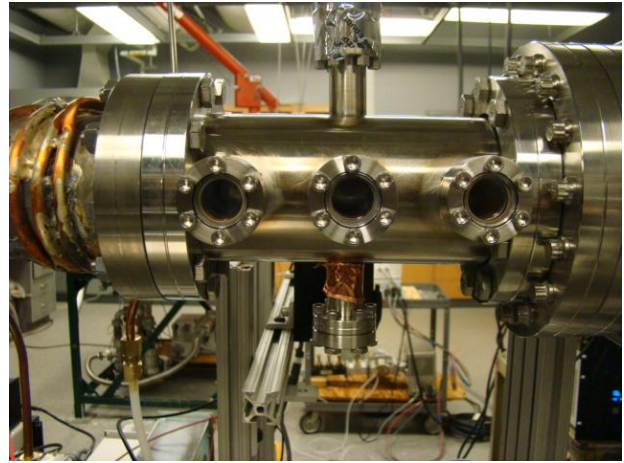

FIG. 2. Photo of cylindrical processing cavity.

The main goal of this preliminary experiment was to see the etching on the outer electrode surface. For this purpose we opted for a variable diameter of the inner electrode with the aim to observe the influence of the surface area of the inner electrode on the plasma properties. A set of variable diameter electrodes with the experimental cylindrical cavity is shown in Fig. 3.

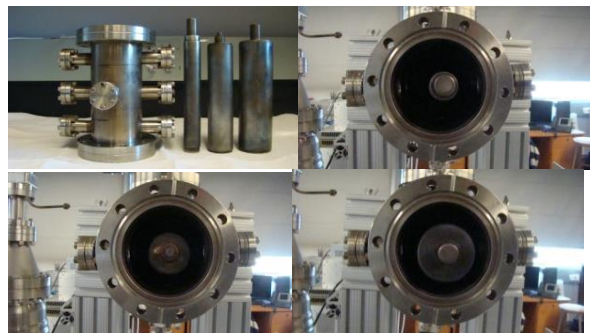

FIG. 3. Cylindrical cavity and inner electrodes with varying diameter. In clockwise direction: the disassembled arrangement and three end on views of the assembled electrode configuration, starting with the lowest diameter of the electrode.

The experimental setup for the cylindrical cavity processing is shown in Fig. 4. The cylindrical chamber is evacuated with the roughing and turbo vacuum pumps. The gas flow and RF power are applied in the opposite direction, shown in Fig. 4. The red arrow is in the gas flow direction and the blue arrow is in the power flow direction. The gas was mixed in a three-branch manifold, each branch containing mass flow meters. The first gas line is attached to the premixed gas cylinder of 15% $Cl_2$ diluted in Ar and the second gas line is connected to the pure Ar cylinder. The third gas line was not used for the present work.

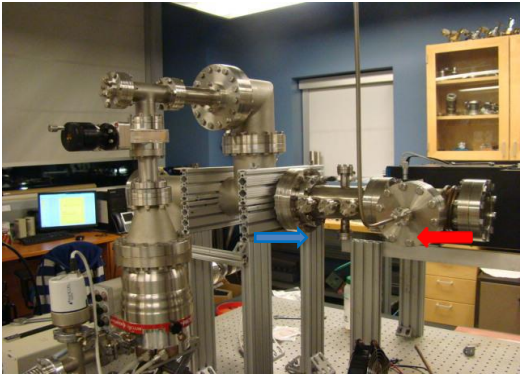

FIG. 4. Photo of experimental setup for cylindrical plasma processing.

We are using a RF power supply with an automatic matching network and an option to connect a dc power supply in series with the RF generator with the purpose to modify the dc bias of the inner electrode. To measure the etch rate of niobium, we opted for ring type samples, shown in Fig. 5. The ring was attached in the full circle to the inside wall of the outer electrode. The ring sample is made of Nb ribbon of 2.5cm width. The diameter of the ring is 7.2cm that correspond to the beam tube diameter of the single cell SRF cavity. It was expected that the ring sample would exhibit a more precise rate of etching performance, as it covered a wide curved surface area, which is several order of magnitude higher than the flat sample area.

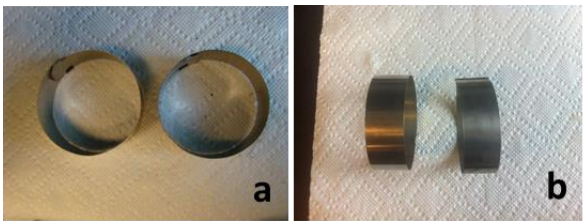

FIG. 5. Photos of etching samples: (a) Top view (b) Side view.

As the plasma properties and in turn the surface processing effect vary substantially with the frequency, pressure, chlorine concentration, temperature and power levels inside the reactor, we have to optimize these parameters for the most efficient and uniform surface material removal from the samples placed on the cavity perimeter.

## IV. EXPERIMENTAL RESULTS

Before presenting the experimental results of the processing study with the cylindrical cavity, shown in Figs. 1 and 3, we will illustrate the normal current effect [6]. Depending on the diameter of the driven electrode there is a certain pressure (above 20 mTorr) where the plasma is completely filling the cavity. When operated below the critical normal pressure the plasma is spread only over a certain section of the cavity. This effect is illustrated by observation at various pressures, shown in Fig. 6.

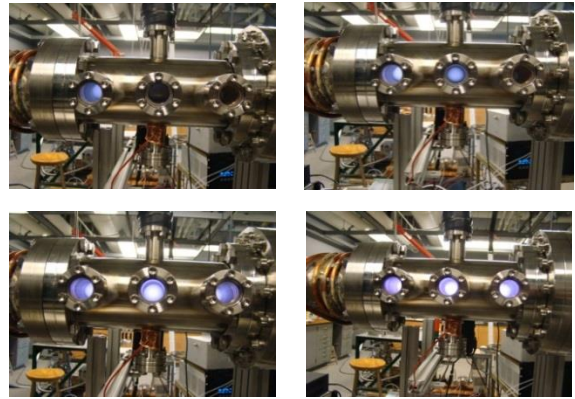

FIG. 6. Spreading of plasma inside the cylindrical cavity at different conditions: top left, lower pressure; top right, medium pressure; bottom left, higher pressure and low power; and bottom right, higher pressure and higher power.

Although much care was taken to keep the plasma spreading in the full cavity, the attempt to etch a niobium sample on the outer wall without applying the positive d.c. bias on the driven electrode was not successful. We tried all possible pressure and power ranges in the given setup, but there was no appreciable material removal to be measured.

### A. Dependence of the etching rate and self-bias voltage on the diameter of the driven electrode

As stated before, when the diameter of the driven electrode is varied, the electrode surface area ratio is changed and, as a consequence, the negative self-bias potential developed across the inner electrode sheath varies. We are, however, interested in the etching rate variation on the grounded (outer) electrode. Constant d.c. bias on the smaller diameter inner electrode does not lift the plasma potential around the grounded electrode by the same value due to its substantially lower surface area.

We have measured the variation of the etching rate with the diameter because the etch rate also depends on the volume of the produced plasma, which increases when we decrease the diameter of the inner electrode and it can affect the etch rate although the plasma potential is lower. All etch rate data are taken at constant d.c. bias of +290 V applied on the inner electrode. The self-bias potential that is negative is measured at the same pressure, power and gas concentration. The measured etching rate and self-bias potential dependence on the inner electrode diameter are given in Fig. 7.

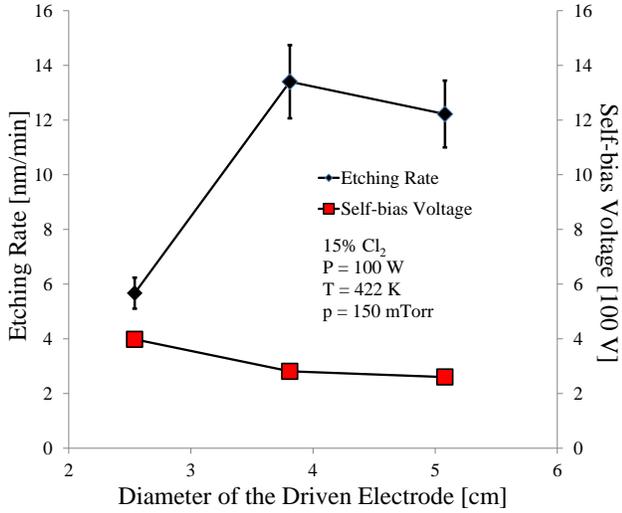

FIG. 7. Etching rate and self-bias dependence on the diameter of the driven electrode. Solid lines are visual guidelines.

## B. Dependence of etching rate on the pressure in the system

There are three competing effects associated with the variation of pressure in the system. Increased pressure means increased concentration of radicals, which are the excited neutral atoms, molecules, and reactive ions of the chlorine. Concentration of molecules is directly proportional to the pressure, but the concentration of reactive ions depends on the RF power and the electronegativity of the discharge. Therefore, at constant RF power, a larger etch rate at higher pressure is indicative of chemical etching, and a smaller etch rate on higher pressure suggests the reactive ion etching mechanism. In addition, pressure increase leads to lower residence time and increase in collision rates among gas molecules which lead to depletion of radicals. Those affect the etching rate as the increase of pressure was obtained by increasing the gas flow rate.

The variation of etching rate with pressure at fixed electrode diameter, constant power, and constant chlorine content is given in Fig. 8. The gas mass flow rates to achieve indicated pressures are 0.25, 0.40, 0.55 and 0.69 l/m, respectively. All measurements were made at T = 422 K. The error in $Cl_2$ concentration was 2%, in power 3 W, in pressure 4 mTorr, in temperature 2K and in dc bias 2V. The etch rate is measured by measuring the mass difference before and after the plasma exposed of the ring sample and dividing it by the area and the density of niobium, and the processing time. The processing time was 100 min in every measurement. The removed layer thickness was 2 to 9 μm.

The calculated error in measurement of mass and area leads to an error in etch rate of 0.82 nm/min. When the error in pressure, power, concentration, temperature and dc bias is included, we estimated the error to be about 10% in etch rate.

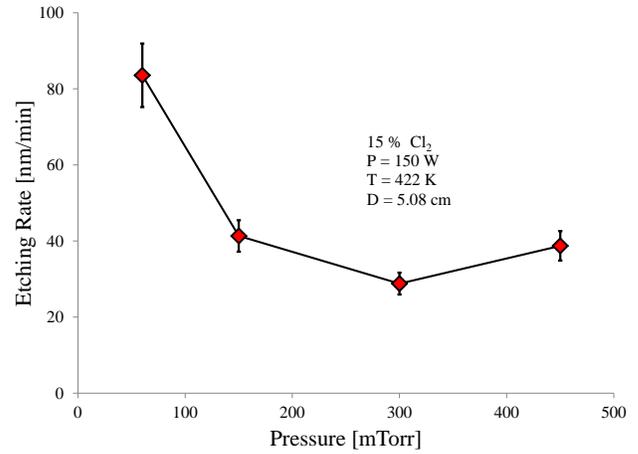

FIG. 8. Etching rate dependence on the gas pressure. Solid line is visual guideline.

This diagram shows that operating at low pressure is more favourable. The maximum etching rate measured was obtained at about 60 mTorr. It then decreased or showed saturation behaviour with the increasing pressure.

## C. Dependence of etching rate on the RF power

The variation of the etching rate with power and all other parameters kept constant is shown in Fig. 9. It follows the general property of reactive RF discharges that the increase in power increases the amount of radicals and ions at a given pressure, which in turn increases the etching rate. In this case the trend did not saturate due to the limited power range.

There is a difference between the present case and in the general trend in planar technology used in the semiconductor etching industry. In planar geometry, increasing power also increases the self-bias of driven electrode which in turn helps in increasing the etch rate. In our case the dc bias is constant and pressure, chlorine concentration and temperature remain constant.

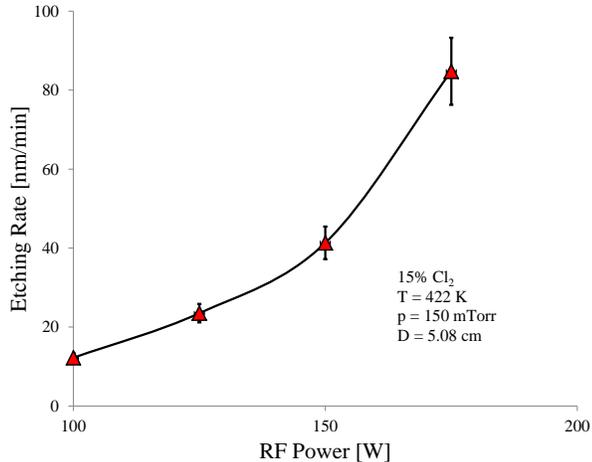

FIG. 9. Etching rate dependence on RF power. Solid line is the exponential fit.

Our data indicate that there are two regimes of the etching rate increase with power. Between 100 and 150 W, the etching rate increases according to the power law with exponent of three. Above 150 W, the increase becomes steeper. This is consistent with the transition to the inductively coupled plasma reported in Refs. [13, 14].

## D. Dependence of etching rate on the chlorine concentration

In Fig. 9 we present the dependence of the average etching rate on $Cl_2$ concentration diluted in Ar. The concentration of chlorine was changed by mixing gases from two cylinders, one 15% $Cl_2$ diluted in Ar and the other pure Ar. The lower percentage of chlorine 10% and 5% was achieved by reducing the flow of $Cl_2$/Ar mixture to 0.26 l/m and 0.13 l/m, respectively, and increasing the flow of pure Ar to achieve the same pressure. In this process, the gas pressure in the experiment was kept constant. Other parameters, such as RF power, d.c. bias, and temperature of substrate remained constant during the experiment.

There are several possible explanations for the saturation effect at relatively low chlorine concentration (see Fig. 10). First, additional chlorine may not have been consumed completely on the surface reactions, which has been observed in Ar/$Cl_2$ discharges [15]. Second, chlorine residence time was not long enough to enable the surface reactions. Further study will elucidate the role of these two effects in the process. We note also that at low power, the discharge is capacitively coupled and electronegative [14], where the electron density is reduced on the expense of negatively charged chlorine ions. This saturation effect may be related to relative electron density reduction due to the increased electronegativity of the discharge. However, positive chlorine ion density, which may be important for the reactive ion etching, should not be affected by the build-up of electronegativity. Consequently, the assumed mechanism of saturation based on the increased electronegativity is not completely certain and has to be studied in more detail.

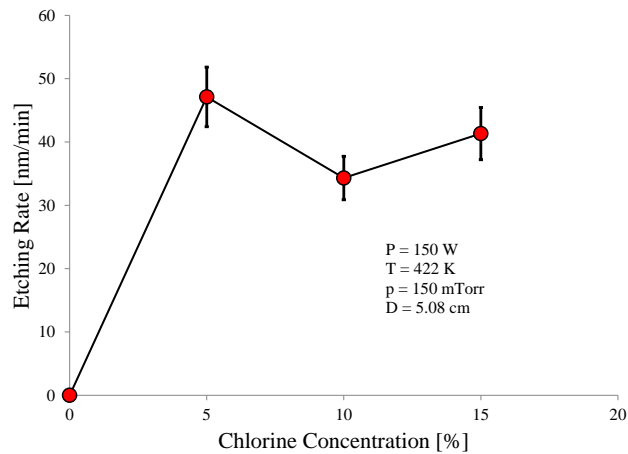

FIG. 10. Etching rate dependence on concentration of chlorine (% Vol) diluted in argon. Solid line is visual guideline.

## V. CONCLUSION

In view of the complex technological challenges facing the development of plasma-assisted SRF surface etching, we have adopted a multistep approach in the transition from flat coupons to the full scale treatment of large cavity surfaces. We have completed the intermediary step of the transition from planar to cylindrical geometry, where a curved Nb sample with the surface area substantially larger than flat coupons was successfully processed at a satisfactory etching rate and at a relatively modest power level. In the present work, we have excited the plasma by a cylindrical coaxial capacitively coupled RF discharge, and demonstrated the effect of its asymmetry by varying the inner electrode diameter and correlating it to the self-bias voltage and the etching rate.

In this work, we have measured the dependence of the etching rate on the driven electrode diameter, RF power, gas pressure, and chlorine concentration. All measurements were performed at 422 K.

The conditions for etching are favorable at large electrode radius, where the self-bias voltage is smaller. We note that the asymmetry is reduced at higher diameter of the driven electrode. This result also means that the sheath voltage amplitude at the grounded electrode is relatively higher, which is the more favorable condition for the reactive ion etching.

Experiments at limited RF power have shown reasonable average etching rates. The increase of the etching rate is expected to somewhat slow down at higher power, but will certainly provide satisfactory etching rates, combined with temperature and pressure adjustments.

Experiments on gas pressure variation indicate favourable conditions at low pressure, as was observed in a number of works in planar technology. It seems that the increase of chlorine concentration above 5% does not provide any substantial benefit to the process.

In the final remark, we have demonstrated a relatively robust etching of a cylindrical Nb sample of diameter comparable to a cylindrical cavity. This result presents a successful intermediary step in the transition from flat coupons etching to the cylindrical cavity wall processing.

## VI. AKNOWLEDGMENT

This work is supported by the Office of High Energy Physics, Office of Science, Department of Energy under Grant No. DE-SC0007879. Thomas Jefferson National Accelerator Facility, Accelerator Division supports J. Upadhyay through fellowship.